\documentclass[aps,prl,a4paper,twocolumn,showpacs,showkeys,preprintnumbers,amsmath,amssymb]{revtex4}
\usepackage{graphicx}
\usepackage{bm}
\usepackage{dcolumn}
\usepackage{color}
\usepackage{pst-plot}
\renewcommand{\texttt}{{}}
\newcommand{\be}{\begin{eqnarray}}
\newcommand{\ee}{\end{eqnarray}}
\usepackage{graphicx}

\begin{document}

\title{Un-spectral dimension and quantum spacetime phases} 

\author{Piero Nicolini}
\thanks{Electronic address: nicolini@th.physik.uni-frankfurt.de}
\affiliation{Frankfurt Institute for Advanced Studies (FIAS),
Institut f\"ur Theoretische Physik, Johann Wolfgang Goethe-Universit\"at, Ruth-Moufang-Strasse 1,
60438 Frankfurt am Main, Germany}

\author{Euro Spallucci}
\thanks{Electronic address: spallucci@trieste.infn.it}
\affiliation{Dipartimento di Fisica, Universit\`a degli Studi di Trieste, 
I.N.F.N. Sezione di Trieste, Strada Costiera 11, 34014 Trieste, Italy}
\date{\small\today}

\begin{abstract} \noindent
In this Letter, we propose a new scenario emerging from the conjectured
presence of a minimal length $\ell$ in the spacetime fabric, on the one side, and
the existence of a new scale invariant, continuous mass spectrum, of
un-particles on the other side.  We
introduce the concept of \textit{un-spectral dimension} $\mathbb{D}_U$      of a 
$d$-dimensional, euclidean (quantum) spacetime, as the spectral dimension
measured by an ``un-particle'' probe. 
We find a  general expression for the un-spectral dimension $\mathbb{D}_U$ labelling
different spacetime phases: a semi-classical phase, where 
ordinary spectral dimension gets contribution from the scaling dimension $d_U$
of the un-particle probe ; a critical "Planckian phase", where four-dimensional spacetime can be effectively considered two-dimensional  when $d_U=1$; a "Trans-Planckian phase", which is accessible to un-particle probes only, where spacetime as we currently understand it looses
its physical meaning.
\end{abstract}
\maketitle
\begin{figure}
 \begin{center}
 \includegraphics[height=4.2cm]{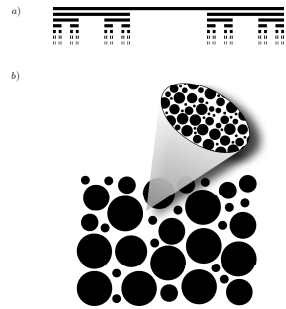}
 \hspace{0.2cm}
      \caption{\label{Plot0}a) A Cantor set. Increasing energy, one moves from the upper continuous surface to the lower fractal. b) A quantum spacetime, showing fractal self-similarity.
 }
 \end{center}
  \end{figure}

If we look at a fractal, for instance the Cantor set in Fig. \ref{Plot0}a), we can grasp the meaning 
of what could be the spacetime in the presence of strong quantum gravity fluctuations.
Indeed fractals capture two of the main features of what we expect to be a quantum spacetime.
When extreme energy/small distance regimes are probed, the spacetime itself changes its own nature 
and exhibits frenzy geometrical and topological fluctuations.  The shorter is the  spacetime scale 
probed, the more involved is the fluctuation pattern. 
Thus, below some fundamental length scale we cannot  model spacetime as a smooth  manifold any longer, 
rather it will look like  a rough and fragmentated (hyper)surface, e.g. a fractal.
Another feature for which a Cantor set, or more generally a fractal, turns out to be quite useful is the 
{\em self-similarity}, namely the property of being exactly similar to a part of itself. In other 
words, fractals are scale invariant because at any magnification there is a smaller piece of the 
fractal that is similar to the whole. Fig. \ref{Plot0}b)  is an artistic representation of 
a fractal spacetime where fractality is represented by a self-similar distribution of holes. From this perspective,  quantum gravity seems to be closely connected to both roughness and scale invariance, 
both features being  supported by 
recent non-perturbative string theory developments like  AdS/CFT duality and $M$-theory.
 A related consideration is given by the (non)renormalizability of gravity following
 from $ \mathrm{mass}^{-2}$ dimension (in natural units) of the Newton constant. On the other hand, in a 
 two-dimensional spacetime the gravitational  
 coupling constant becomes dimensionless and gravity is expected to be power-counting renormalizable. 
 This special feature is accompanied by the fact that in two dimensions the spacetime is conformally 
 flat and field theories more naturally enjoy properties like conformal invariance. In support of this 
 line of reasoning there is the concept of spectral dimension, one of the most intriguing features of 
 a quantum spacetime. If we expect an increasing degree of fuzziness in the quantum regime, then
 we must accept the idea  that also spacetime dimension should be reviewed.  As
 the classical manifold dissolves into a sort of fractal dust, the very concept of ``dimensionality'' must change from an assigned property into a dynamical quantity running with the energy scale of the probe.
 An effective way to measure the actual dimension of a quantum manifold consists in studying 
 the diffusion of a test particle. The dynamics of the process is encoded into the heat kernel equation
\begin{eqnarray}
\Delta\, K\left(\, x\ ,y\ ; s\,\right)=
\frac{\partial}{\partial s}\,
K\left(\, x\ ,y\ ; s\,\right)
\end{eqnarray}
where $s$ is a fictitious diffusion time of dimension of a length squared, $\Delta$ is the Laplace 
operator and $K\left(\, x\ ,y\ ; s\,\right)$ is the heat kernel, representing the probability density 
of diffusion from $x$ to $y$ in a ``lapse of time $s$''. The initial condition for the diffusion
process is that the test particle starts  from $x$ at $s=0$
\be
K\left(\, x\ ,y\ ; 0\,\right)=\frac{\delta^d(x-y)}{\sqrt{\det g_{ab}}} 
\label{ic}
\ee
where $\delta^d(x-y)$ is the $d$-dimensional Dirac delta, $d$ is a integer number representing the 
topological dimension  and $g_{ab}$ is the metric of the manifold. If we consider a closed
random path, i.e. $x=y$, we can define the \textit{return probability} by integrating the
Kernel over all spacetime and factorizing out the total invariant volume
\be
P(s)=\frac{\int d^d x\sqrt{{\mathrm \det} \ g_{ab}}\ K(x,x;s)}{\int d^d x\sqrt{{\mathrm \det}\ g_{ab}}}.
\ee
From $P(s)$ we can define the \textit{spectral dimension} as
\be
{\mathbb D}=-2 \, \frac{\partial \ln P_g(s)}{\partial \ln s}\ .
\label{spectrad}
\ee
It is easy to show that in flat space, for a ``free'' diffusion, the return probability is 
$P(s)=(4\pi s)^{ -d/2}$ and the spectral dimension is ${\mathbb D}=d$. In the presence of gravity, 
the above formula can be  yet employed to check an effective dimensional reduction, even if the 
large $s$ limit holds only on local patches of the manifold which approximates the tangent space. 
The importance of the spectral dimension lies in the fact that it could provide a glimpse about a 
crucial feature of a quantum manifold: if it turned out that in the quantum gravity regime the actual 
dimension measured by the diffusion process is two, we could conclude that gravity is a 
renormalizable theory, overcoming the conventional difficulties about its quantization. As a result 
there have been many attempts to calculate the spectral dimension  
 \cite{Carlip:2009kf} and it has been found that $\mathbb{D}$ tends to the value $2$ for scales 
 approaching $\ell$, an effective minimal length in the manifold \cite{Coherent,Modesto:2009qc}.
However, in all the approaches above it is understood that short distance can be
probed only by ultra-relativistic objects with a negligible rest mass. Thus, scale invariance
is kinematically realized in a light-cone type limit. Indeed, if we consider the heat equation
for a massive particle we find
\begin{eqnarray}
\tilde\Delta\, K\left(\, x\ ,y\ ; s\,\right)=
\frac{\partial}{\partial s}\,
K\left(\, x\ ,y\ ; s\,\right)
\label{massive}
\end{eqnarray}
where the operator $\tilde\Delta=\Delta-m^2$ includes a non-differential term $m^2$. From the
definition (\ref{spectrad}) we get
\be
{\mathbb D}=-2s\frac{\int d^d x\sqrt{{\mathrm \det} \ g_{ab}}\ \Delta K(x,x;s)}{\int d^d x
\sqrt{{\mathrm \det} \ g_{ab}}\  K(x,x;s)}+2sm^2. \label{md}
\ee
The first term in the r.h.s of (\ref{md}) leads to a constant, i.e. scale independent, value of the spectral dimension,
 but the second one is linear in $s$ and diverges for asymptotic diffusion times  spoiling 
 a meaningful  definition of $\mathbb{D}$. From this viewpoint, one concludes that the spectral 
 dimension can be safely introduced only in a scale invariant framework. In other words, 
 spacetime spectral dimension cannot be probed by massive objects as they break scale invariance.

 In this Letter we are going to present a new, scale invariant, procedure to compute $\mathbb D$
 by means of a massive probe providing a non-trivial modification to the standard definition.
 It may sound odd to preserve scale invariance in the presence of a massive object, but
this problem can be by-passed by using \textit{un-particle probes} borrowed from a recently
proposed extension of the elementary particle standard model. 
The  new idea is that there exists a new high-energy sector of the
particle standard model where the fundamental objects display a continuous, 
scale-invariant, mass spectrum in alternative to the discrete mass spectrum
of ordinary elementary particles. This new ``stuff'' is very weakly coupled 
to ordinary matter below some threshold energy, say some TeV  \cite{Georgi:2007ek}. Beyond these energies
the standard model fields interact with a new scale invariant sector described by the so-called
Banks-Zaks (BZ) fields. The interaction is mediated by very heavy particles of mass $M_U$. 
Below $M_U$, the interaction leads to non-renormalizable effective couplings of 
the form $ O_{sm}O_{BZ}/M_U^k$ where $O_{sm}$ is an operator with mass dimension $d_{sm}$ built out 
of the standard model fields and $O_{BZ}$ is the equivalent for BZ fields. 
 The scale invariance properties of the BZ sector emerge below a scale $\Lambda_U$, through
 dimensional transmutation of the BZ fields into un-particle fields.
The above interaction term  becomes $C_U(\Lambda_U)^{d_{BZ}-d_U} O_{sm}O_{U}/M_U^k$,
where the BZ operators $O_{BZ}$  match un-particle operators $O_{U}$; $d_U$ is the un-particle 
scaling dimension and $C_U$ is a normalization constant. In this scenario the BZ fields decouple from 
ordinary matter at low energies and therefore the interaction $ O_{sm}O_{BZ}/M_U^k$ should not affect 
the scale invariant properties of un-particles. 
Even if the scaling dimension can be arbitrarily large, it is customary to assume that $1<d_U<2$. 
  Recently it has been argued that un-particle might affect the gravitational interaction too: 
 indeed un-gravity could arise from the un-graviton exchange among massive particles
\cite{Goldberg:2007tt}. In addition by exploiting results coming from Cavendish experiments one obtains that eventual un-particle corrections to Newton's law might occur at energy scales higher than TeV, in agreement with the basic hypotheses about un-particle physics. Following this line of reasoning, un-gravity corrections to
 Schwarzschild metric have been perturbatively derived in \cite{Mureika:2007nc} and confirmed
at the non-perturbative level  by solving the field equations derived from
an effective action including the un-graviton corrections at all order \cite{Gaete:2010ip}. As a special 
result, the Hawking temperature and the Bekenstein entropy of the un-Schwarzschild black hole 
suggest that the dimension of the horizon is a non-integer number $2d_U$.
These examples suggest that un-particle physics provides a tool to implement fractalization of 
conventional scenarios,  by forcing the presence of the scaling dimension $d_U$. 
From this vantage point, it is almost compelling  to explore the spectral dimension,
i.e. the fractal structure of the Planckian spacetime,  by means of
un-particle probes. For sake of clarity we stress that in this case un-particles do not generate gravity since we are considering a generic manifold either classical or quantum. 
In the simplest case of a scalar un-particle the Green function 
turns out to be \cite{Gaete:2008wg}
\be
&& G_U( x-y )=A_{d_U}\int_0^\infty dm^2 (m^2)^{d_U-2} G( x-y; m^2 )\nonumber
\ee
where $d_U$ controls a continuous mass spectrum, while
\be
A_{d_U}=\frac{8\pi^{5/2}}{(\ \Lambda_U^2\ )^{d_U-1} (\ 2\pi \ )^{d_U}}
\frac{\Gamma(\ d_U+1/2\ )}{\Gamma(\ d_U-1\ )\ \Gamma(\ 2d_U\ )}.\nonumber
\ee
The heat kernel $K_U(\ x, y;\ s\ )$ can be obtained from 
\be
&&G_U(\ x-y\ )=\int_0^\infty ds\ K_U(\ x, y;\ s\ )\\
&&=A_{d_U}\int_0^\infty ds\ \int_0^\infty dm^2\ (m^2)^{d_U-2}\ K(x, y;\ s )
\label{ungreen}\nonumber
\ee
where $K(x, y;\ s )$ is the solution of Eq. (\ref{massive}).
We can study the diffusion of an un-field 
\begin{eqnarray}
\Delta_U\, K_U\left(\, x\ ,y\ ; s\,\right)=
\frac{\partial}{\partial s}\,
K_U\left(\, x\ ,y\ ; s\,\right)
\label{undiffusion}
\end{eqnarray}
where the un-Laplacian acquires an extra term depending on the un-particle sector 
$\Delta_U=\Delta-(d_U-1)/s$.
Eq. (\ref{undiffusion}) can be classified as an inhomogeneous heat equation, whose initial 
conditions are like in (\ref{ic}).
 Employing a one-dimensional heat conduction analogy, we could say that our problem is equivalent 
 to that of a bar which is subjected to a time dependent ``heat source'' $(d_U-1)/s$. 
 In other words, the heat released by the un-particle term is spatially uniform along the length of 
 the bar and the scale invariance is preserved. We notice that for $d_U=1$ the above equation 
 becomes the homogeneous heat equation, in agreement with the fact that un-particle corrections 
 vanish for $d_U=1$ as in Ref. \cite{Goldberg:2007tt}. Therefore, from Eq. (\ref{spectrad}), 
 we can define  the un-spectral dimension as 
\be
{\mathbb D}_U=-2s\frac{\int d^d x\sqrt{{\mathrm \det} \ g_{ab}}\ 
\Delta K_U(x,x;s)}{\int d^d x\sqrt{{\mathrm \det} \ g_{ab}}\  K_U(x,x;s)}+ 
\frac{2\Gamma( d_U )}{\Gamma( d_U-1 )}\ .\label{und}
\ee
The above formula can be manipulated to obtain 
\begin{equation}
{\mathbb D}_U={\mathbb D}+2d_U-2\ , \label{main}
\end{equation}
Eq.(\ref{main}) is the main result of this work. In analogy with the Hausdorff dimension (see Ref. \cite{Hausdorff}),  
we see that in the chosen range for the scale dimension  ${\mathbb D}_U\ge {\mathbb D}$,
 while ${\mathbb D}_U= {\mathbb D}$ for $d_U=1$ only. 
This increase of the dimension measured by the diffusion process can be explained in terms of the 
presence of an additional sector, i.e. the un-particles, with respect to the conventional standard 
model fields calculation. On the other hand, for the specific case ${\mathbb D}=2$, one finds that the 
un-spectral dimension depends uniquely on $d_U$ and is ${\mathbb D}_U=2d_U$. As a result for a 
diffusion process in a flat plane, $d=2$ and 
\be
&&K_{p}\left(\, x\ ,x\ ; s\,\right)= A_{d_U}\int_0^\infty dm^2
\ (m^2)^{d_U-2}\ \frac{e^{-m^2 s}}{(\ 4\pi s\ )^{d/2}} \nonumber
\ee
we obtain ${\mathbb D}=d=2$ and ${\mathbb D}_U=2d_U$. As a first application of Eq.(\ref{main})
we are going to investigate 
the nature of the un-Schwarzschild 
black hole horizon, but we need to do a little step forward. The horizon is a curved surface and 
therefore the diffusion must take into account this effect through non-trivial Seeley-deWitt 
coefficients in the heat kernel representation. A further modification occurs in the Laplace 
operator which acquires a non-minimal coupling to the Ricci scalar to preserve scale invariance
$\Delta\longrightarrow \Delta_g\equiv \Delta -\xi_d \, R$ with $\xi_d\equiv (1/4)(d-2)/(d-1)$.
As a result the heat kernel in the presence of gravity reads 
\begin{eqnarray}
&&K_{g}\left( x , x ; s \right)=A_{d_U}\times\\&&\int_0^\infty dm^2 (m^2)^{d_U-2} 
\frac{e^{-m^2 s}}{( 4\pi s )^{d/2}}\left[a_0  + \sum_{n=1}^\infty\, s^n   a_n\left(
x ,x \right)\right].\nonumber
\end{eqnarray}
Since the un-particle ``heat source'' preserves scale invariance and does not affect the manifold 
coordinate $x$, its contribution $2\Gamma( d_U )/\Gamma( d_U-1 )$ will be unchanged in the presence 
of gravity. Here, for the sake of clarity, we provide only the gravity primary corrections
\be
{\mathbb D}_U=d+2d_U-2-2s\frac{\int d^d x\sqrt{{\mathrm \det}  g_{ab}} 
\left[a_1+2a_2s+\dots\right]}{\int d^d x\sqrt{{\mathrm \det}  g_{ab}} 
\left[a_0+a_1s+\dots\right]}.&&\nonumber
\ee
We remind that the above formula holds for small diffusion times only. 
Indeed for a generic topological dimension gravity introduces a scale in the conventional term for 
the spectral dimension. This is the reason why there is a breaking of the scale invariance analogous 
to the introduction of a mass as in (\ref{massive}). This is not the case for $d=2$. Indeed, when
one considers the un-Schwarzschild black hole horizon, 
we have a conformal invariant diffusion, propagating on a conformally flat manifold. Thus, the 
Green function (\ref{ungreen}) reduces to the flat space un-particle Green function and we can 
conclude that ${\mathbb D}_U=2d_U$, indicating ``fractalization'' of the surface.
This would confirm the argument in \cite{Gaete:2010ip} according to which the un-Schwarzschild 
horizon is exactly a $2d_U$-dimensional fractal surface built up by  un-gravitons trapped at the
Schwarzschild radius.

Up to now, we have considered the case of ``classical'' background manifold in the sense that two
points (events) can be arbitrarily closed. In other words, we have not  considered the intrinsic uncertainty
in the localization of a single point when it is left free to fluctuate quantum mechanically.
Since our model  of  "quantum manifold"  would like to account  for
  both a short-distance increasing loss of resolution and self similarity, it is compelling to understand how the un-spectral dimension behaves in regards of both properties. To this purpose, we implement the graininess in spacetime along the lines of 
Ref. \cite{Modesto:2009qc} by studying a diffusion process governed by the same heat equation as 
in (\ref{undiffusion}), but with a modified initial condition 
$K_{\ell}\left( x ,y ; 0 \right)=\frac{\rho_\ell\left( x ,y \right)}{\sqrt{\det g_{ab}}}$ .
$\rho_\ell\left( x ,y \right)$ is  a Gaussian distribution replacing the former Dirac-delta.
The width $\ell$  is the minimal uncertainty in the distance between two
fluctuating points, or the best resolution which is compatible with the quantum nature of the background manifold. This loss of resolution primarily affects the early, short-distance, stages of the diffusion process, while at distance large with respect to $\ell$, the  diffusion is insensitive to the graininess of the manifold.
From a thermal point of view,
the manifold behaves like a "boiling surface", whose thermal instability 
sustains the Gaussian profile  preventing it from collapsing into a Dirac delta. 
In what follows,  the role of fluctuations in the Riemannian curvature, which is a geometrical attribute of
a classical, smooth, manifold becomes less and less relevant with respect to the graininess of the manifold itself. Thus, for our next purpose it is enough to consider the flat, but accounting for quantum uncertainty , heat kernel 
\begin{eqnarray}
K_{\ell}\left( x , y ; s \right)=A_{d_U}\int_0^\infty dm^2\ (m^2)^{d_U-2}\  
\frac{e^{-m^2 s}e^{- \frac{\left( x-y \right)^2}{4 (s + \ell^2)
 } }}{\left[  4\pi \left(s+\ell^2 \right) \right]^{d/2} }.\nonumber
\end{eqnarray}
The resulting un-spectral dimension  turns out to be 
\be
{\mathbb D}_U=\frac{s}{s+\ell^2}\ d-2+2d_U.
\label{unspectrall}
\ee
Eq.(\ref{unspectrall}) is the second main result of this work. It provides a new physical interpretation
of the fundamental constant $ \ell $ as the \textit{transition scale} between  different
phases of the background spacetime.  
Long random walks, where $s>> \ell^2$ , 
test a semi-classical manifold characterized by ${\mathbb D}_U =  d-2 + 2d_U $. For the
special case $d=2$ the manifold un-spectral dimension is totally determined by the scaling
parameter $d_U$, as in the case of the un-Schwarzschild black hole. Conversely, for $d_U=1$
the un-matter effects decouple and the un-spectral dimension matches the topological dimension $d$.
In the \textit{critical}, say Planckian, regime we have $s\approx \ell^2$ and ${\mathbb D}_U =  2d_U-2 +d/2$.
For $d=4$, we see that at Planck scale  spacetime dimension is totally determined by the scaling dimension, as it is ${\mathbb D}_U =  2d_U $,
just like in the case of the un-Schwarzschild black hole. In the particular case $d_U=1$, we obtain  
the dynamical reduction to ${\mathbb D}_U =  2$ necessary to get a power counting renormalizable
quantum theory of gravity. However, this is not the end of story. 
Un-matter  allows us to access a new "trans-Planckian" phase which cannot be probed by 
any sort of ordinary matter.
Short paths, where $s<< \ell^2$  measure ${\mathbb D}_U =  2d_U -2 +O\left(\, s/l^2\,\right)$ which is
non-negative only in virtue of the scaling dimension $d_U\ge 1$.  We see that for $ d_U< 2 $ 
the un-particle probe scatters across something which we can dub  "spacetime vapor"  to be consistent
with the thermal interpretation of the diffusion process. Moreover, as $d_U\to 1$ then ${\mathbb D}_U\to 0$ leading to the ultimate disintegration of space and time as we understand them.
This new picture follows from the introduction of   un-spectral dimension, 
as a dimension measured by a scale invariant continuous mass spectrum probe.
It may be worth to remark that even in the 
``worst-case-scenario'', where un-particles were not found at LHC as physical
objects, the definition (\ref{und}) would provide an alternative realization of
a scale invariant diffusion process, not advocating a light-cone limit. 
The ``extra-bonus'' of this approach is to bring into the definition of un-spectral dimension
the real parameter $d_U$ leading to a clear fractalization of the background space(time) and
the appearance of a new phase which is forbidden to standard matter probes.
As both  spectral and  Hausdorff dimensions are employed in a variety of diffusion problems,
we expect the un-spectral dimension  to have an equivalent impact in frameworks different from the
one considered in the present Letter. 
\begin{acknowledgments}
\noindent P.N. is supported by the Helmholtz International Center for FAIR within the
framework of the LOEWE program (Landesoffensive zur Entwicklung Wissenschaftlich-\"{O}konomischer Exzellenz) launched by the State of Hesse. 
\end{acknowledgments}


\begin{thebibliography}{99}
\bibitem{Carlip:2009kf}
  S.~Carlip,
  arXiv:0909.3329 [gr-qc].

  J.~Ambjorn, J.~Jurkiewicz and R.~Loll,
  Phys.\ Rev.\ Lett.\  {\bf 95}, 171301 (2005).

  L.~Modesto,
  arXiv:0812.2214 [gr-qc].
  
  L.~Modesto,
  arXiv:0905.1665 [gr-qc].

  
  F.~Caravelli and L.~Modesto,
  arXiv:0905.2170 [gr-qc].

  E.~Magliaro, C.~Perini and L.~Modesto,
  arXiv:0911.0437 [gr-qc].


  O.~Lauscher and M.~Reuter,
  JHEP {\bf 0510}, 050 (2005).
  
  D.~Benedetti,
  Phys.\ Rev.\ Lett.\  {\bf 102}, 111303 (2009).
  


\bibitem{Coherent}
  A. Smailagic and E. Spallucci,
 J. Phys. A{\bf 36}, L467 (2003).


  A. Smailagic and E. Spallucci,
  J. Phys. A{\bf 36}, L517 (2003).

  A.~Smailagic and E.~Spallucci,
  J. Phys. A  {\bf 37}, 1 (2004)
  [Erratum-ibid.  A {\bf 37}, 7169 (2004)].

  E.~Spallucci, A.~Smailagic and P.~Nicolini,
  Phys. Rev.  D {\bf 73}, 084004 (2006).




  P.~Nicolini, A.~Smailagic and E.~Spallucci,
ESA Spec. Publ. \textbf{637}, 11.1 (2006). 

  P.~Nicolini,
  J. Phys. A  {\bf 38}, L631 (2005).

   P.~Nicolini, A.~Smailagic and E.~Spallucci,
  Phys. Lett.  B {\bf 632}, 547 (2006).
  
  T.~G.~Rizzo,
  JHEP {\bf 0609}, 021 (2006).
  
  S.~Ansoldi, P.~Nicolini, A.~Smailagic and E.~Spallucci,
  Phys. Lett.  B {\bf 645}, 261 (2007).
  
  E.~Spallucci, A.~Smailagic and P.~Nicolini,
  Phys. Lett.  B {\bf 670}, 449 (2009).
  
  P.~Nicolini and E.~Spallucci,
    Class.\ Quant.\ Grav.\  {\bf 27}, 015010 (2010).
  
  R.~Casadio and P.~Nicolini,
  JHEP {\bf 0811}, 072 (2008).
  
  P.~Nicolini,
  Int. J. Mod. Phys.  A {\bf 24}, 1229 (2009).

  M.~Rinaldi,
  arXiv:0908.1949 [gr-qc].
  
  P.~Nicolini and M.~Rinaldi,
  arXiv:0910.2860 [hep-th].

   D.~Batic and P.~Nicolini,
  Phys.\ Lett.\  B {\bf 692}, 32 (2010).


   M.~Bleicher and P.~Nicolini,
  J.\ Phys.\ Conf.\ Ser.\  {\bf 237}, 012008 (2010).


A.~Smailagic and E.~Spallucci,
  Phys.\ Lett.\  B {\bf 688}, 82 (2010).

  
\bibitem{Modesto:2009qc}
  L.~Modesto and P.~Nicolini,
  Phys.\ Rev.\  D {\bf 81}, 104040 (2010).

\bibitem{Georgi:2007ek}
  H.~Georgi,
  Phys.\ Rev.\ Lett.\  {\bf 98}, 221601 (2007).
  
\bibitem{Goldberg:2007tt}
  H.~Goldberg and P.~Nath,
  Phys.\ Rev.\ Lett.\  {\bf 100}, 031803 (2008).

\bibitem{Mureika:2007nc}
  J.~R.~Mureika,
  Phys.\ Lett.\  B {\bf 660}, 561 (2008).
  
  J.~R.~Mureika,
  Phys.\ Rev.\  D {\bf 79}, 056003 (2009).


\bibitem{Gaete:2010ip}
 P.~Gaete, J.~A.~Helayel-Neto and E.~Spallucci,
  Phys.\ Lett.\  B {\bf 693}, 155 (2010).


\bibitem{Gaete:2008wg}
  P.~Gaete and E.~Spallucci,
  Phys.\ Lett.\  B {\bf 661}, 319 (2008).
  
\bibitem{Hausdorff}
  S.~Ansoldi, A.~Aurilia and E.~Spallucci,
  Phys.\ Rev.\  D {\bf 56}, 2352 (1997).

  S.~Ansoldi, A.~Aurilia and E.~Spallucci,
  Chaos Solitons Fractals {\bf 10}, 197 (1999).

  A.~Aurilia, S.~Ansoldi and E.~Spallucci,
  Class.\ Quant.\ Grav.\  {\bf 19}, 3207 (2002).


\end{thebibliography}
\end{document}